\documentclass[pdflatex,sn-mathphys-num]{sn-jnl}

\usepackage{graphicx}%
\usepackage{multirow}%
\usepackage{amsmath,amssymb,amsfonts}%
\usepackage{amsthm}%
\usepackage{mathrsfs}%
\usepackage[title]{appendix}%
\usepackage{xcolor}%
\usepackage{textcomp}%
\usepackage{manyfoot}%
\usepackage{booktabs}%
\usepackage{pifont}

\theoremstyle{thmstyleone}%
\newtheorem{theorem}{Theorem}
\theoremstyle{thmstyletwo}%

\theoremstyle{thmstylethree}%

\newcommand{\xm}{\ding{51}}%
\newcommand{\cm}{\ding{52}}%

\newtheorem{ax}{A}

\newtheorem{corollary}{Corollary}

\newtheorem{lemma}{Lemma}

\newcommand{\epr}{\hfill $\Box$\mbox{}\\ }

\newcommand{\SYM}{Symmetry}
\newcommand{\Unit}{Unit}
\newcommand{\LB}{Lower Bound}
\newcommand{\TR}{Transfer}

\newcommand{\ERJ}{Expansion Responsiveness-J}
\newcommand{\ERH}{Expansion Responsiveness-H}
\newcommand{\ERO}{Expansion Responsiveness-O}
\newcommand{\ERJsh}{Exp. Resp.-J}
\newcommand{\ERHsh}{Exp. Resp.-H}
\newcommand{\EROsh}{Exp. Resp.-O}
\newcommand{\StTSES}{Constant Sensitivity-O}

\newcommand{\StTSESS}{Constant Sensitivity-S}
\newcommand{\TSES}{Constant Sensitivity-O}

\newcommand{\TSESS}{Constant Sensitivity-S}
\newcommand{\StTSESsh}{Const. Sens.-O}

\newcommand{\StTSESSsh}{Const. Sens.-S}

\newcommand{\OES}{One Empty Set}
\newcommand{\TES}{Two Empty Sets}
\newcommand{\EI}{Expansion Invariance}
\newcommand{\EIsh}{Exp. Inv.}

\newcommand{\GADD}{General Additivity}

\newcommand{\RI}{Replication Invariance}
\newcommand{\StRI}{Replication Invariance}
\newcommand{\StRIsh}{Rep. Inv.}
\newcommand{\NEU}{Neutrality}
\newcommand{\IND}{Independence}
\newcommand{\ADD}{Additivity}
\newcommand{\SA}{Super-Additivity}

\newcommand{\rb}[1]{\rotatebox{90}{#1}}
\newcommand{\NN}{\mathbb{N}}
\newcommand{\RR}{\mathbb{R}}
\newcommand{\QQ}{\mathbb{Q}}

\let\emptyset\varnothing

\newcommand{\bpr}{\noindent {\bf Proof.} \hspace{1 em}}

\providecommand{\card}[1]{\left\lvert #1 \right\rvert}

\newcommand{\parttwo}{(Part 2)} 
\newcommand{\partone}{(Part 1)} 

\newcounter{myindex}

\begin{document}

\title[Article Title]{Axiomatic characterizations of dissimilarity orderings and distances between sets}

\author*[1]{\fnm{Thierry} \sur{Marchant}}\email{thierry.marchant@ugent.be}

\author[2]{\fnm{Sandip} \sur{Sarkar}}\email{sandips@goa.bits-pilani.ac.in}
\equalcont{Authors are listed alphabetically. They have contributed equally. }

\affil*[1]{ \orgname{Ghent University}, \country{Belgium}}

\affil[2]{\orgdiv{Department of Economics and Finance}, \orgname{Birla Institute of Technology and Science, Pilani}, \orgaddress{\street{K K Birla Goa Campus, Zuarinagar}, \city{Sancoale}, \postcode{403726}, \state{Goa}, \country{India}}}

\abstract{We axiomatically characterize the orderings of pairs of sets induced by several distances: Hamming, Jaccard,  S\o rensen-Dice  and  Overlap. We also axiomatically characterize these distances. Our axioms are  properties describing how a distance changes when we perform elementary modifications of the sets, like adding one element to one of the sets, to both sets, swapping both sets, permuting some elements, etc.}


\maketitle

\section{Introduction}

Researchers  are often interested in quantifying the dissimilarity (or similarity\footnote{In the present paper, we focus on dissimilarity. The transition from one to the other is often simple, assuming that similarity $= 1 -$dissimilarity.}) between two sets, across a variety of fields, including classification \citep{Kartaletal2026,PerisicVanbelle2025}, decision making \citep{ZhaoZhangXiao2024},  network analysis \citep{WangChen2025}, misinformation detection \citep{HouOfoghiYearwood2026}, chemistry \citep{Maggiora2014}, information retrieval \citep{bookstein2002generalized}, graph compression \citep{GiancottiGulloGuzziSerraVeltri2026}, scientometrics \citep{Baccinietal2024}, and  genomics \citep{MajiShahPaul2017}, among others. Although numerous measures exist to quantify the dissimilarity between two  sets, our focus will be on  four widely used distances: the Jaccard distance, the S\o rensen-Dice distance, the Hamming distance, and the Overlap distance \citep{DezaDeza2009}. 

In some applications, users of distances are actually not interested in the distance itself, but rather in the ordering induced by the distance. In other words, it may be important to know whether the dissimilarity between sets $A$ and $B$ is larger than that between $C$ and $D$, while the numerical value of the distance between $A$ and $B$ (or $C$ and $D$) does not matter. An insightful illustration of preferring orderings over distances in the context of image retrieval is provided by \cite{OmhoverRifqiDetyniecki2006}. They highlight that image retrieval systems often output a list of images ordered according to the similarity between a description of the image and the description corresponding to the query, without even displaying the corresponding distances. In other applications, the numerical value of the distance is of interest. For instance, in information theory, the Hamming distance is often multiplied by some probability (transition probability or rate of the error correction code) and this product is further used in other calculations \citep{ChenWornell1999}.

Given that the four above-mentioned distances or orderings are widely applied, it is important to  be aware of their properties. This can help make an informed choice of a distance  for a specific application. Once a distance is chosen, it can also help use and interpret it in a meaningful way. Our primary objective in this paper is therefore to axiomatically characterize the four distances as well as the corresponding orderings.\footnote{Readers interested in the role and importance of axiomatic characterizations will find a deep and extensive treatment of these questions in \citet[Ch.21]{KLST90}. Less abstract  and more concise answers are given in \cite{Thomson2001,GilboaPostlewaiteSamuelsonSchmeidler2019}.}

As far as we have surveyed, we did not come across any study axiomatically characterizing one of the four chosen  orderings. We also did not find any axiomatic characterization of the distances, with the only exception being the axiomatic characterization of the Jaccard distance by \cite{Gerasimou2024}. For this characterization, \citeauthor{Gerasimou2024} uses a very strong axiom named constant marginal sensitivity, which requires that, if any element belonging to two distinct sets is removed from one but not both sets, then the dissimilarity increases by the inverse of the number of elements in the union of the two sets. We provide three alternative characterizations of the Jaccard distance, one with a much weaker version of the constant sensitivity condition, and another one by replacing the constant sensitivity condition by several weak conditions and  additivity. The third characterization is obtained using the triangle inequality. We adopt a similar approach for the characterization of the Hamming distance. For the S\o rensen-Dice and the Overlap distance, we establish that the distance does not satisfy any kind of additivity and we therefore rely only on some notions of constant sensitivity to characterize the corresponding distances.

The rest of the paper is organized as follows. In section \ref{sec.formal.frame}, we discuss the formal framework for our analysis. We characterize the dissimilarity orderings and distances in sections \ref{sec.orderings} and \ref{sec.distances}, respectively. A general discussion is  presented in section \ref{sec.discussion}. The proofs of the Theorems are placed in section \ref{sec.proofs}. 


\section{Notation and definitions}
\label{sec.formal.frame}

Let $\NN, \QQ, \RR$ respectively denote the natural numbers (positive integers), the rational numbers and the real numbers. 
Let $X$ be a set and $Y$ the set of all finite subsets of $X$.  In most results, we will impose that $X$ be infinite (countable or not).
For definitions of distances, metrics, semi-metrics, and so on, we  follow \cite{DezaDeza2009}.
A distance is a mapping $I: Y\times Y \to \RR$ such that, for all $A,B \in Y$,
\begin{itemize}
\item $I(A,B) \geq 0$,
\item $I(A,B)=I(B,A)$,
\item $I(A,A)=0$.
\end{itemize}
A distance is a near-metric if it satisfies the weak triangle inequality
\begin{equation}
\label{eq.weak.triangle}
I(A, B) \leq  \gamma \big( I(A, C) + I(C, B) \big),
\end{equation}
for some $\gamma \geq 1$.
A distance is a semi-metric if it satisfies \eqref{eq.weak.triangle} with $\gamma=1$ and
a semi-metric is a metric if it also satisfies, for all $A,B \in Y$, $I(A,B)=0$ iff $A=B$. 

A dissimilarity ordering (or ordering for short) is a binary relation on $Y\times Y$ satisfying, for all $A,B,C,D,E,F \in Y$, (1) $(A,B) \succsim (C,D)$ or $(C,D) \succsim (A,B)$ (completeness) and (2) $(A,B) \succsim (C,D) \wedge (C,D) \succsim (E,F) \Rightarrow (A,B) \succsim (E,F)$ (transitivity). The statement $(A,B) \succsim (C,D)$ is interpreted as `the dissimilarity between $A$ and $B$ is not smaller than between $C$ and $D$.' The asymmetric and symmetric parts of $\succsim$ are denoted by $\succ$ and $\sim$; they are defined as usual.

Any distance $I$ induces a dissimilarity ordering $\succsim_{I}$ defined by 
\begin{equation}
\label{num.repr}
(A,B) \succsim_{I} (C,D) \iff I(A,B) \geq I(C,D),
\end{equation}
for all $A,B,C,D \in Y$. 
 Notice however that there is no one-to-one correspondence between distances and orderings. In particular, infinitely many distances correspond to a single ordering. For instance, the distances $J, J^{2}, \sqrt{J}, \exp(J)$ and $\sin(J)$ all induce the same ordering $\succsim_{J}$. In general, all strictly increasing transformations of a distance $I$ induce the same ordering $\succsim_{I}$.
Moreover, some orderings are not induced by a distance. For instance, let $X=\RR$ and define the  ordering $\succsim_{L}$  by (i) $(A,B) \sim_{L} (C,D)$ iff $\sup(A)=\sup(C)$ and $\sup(B)=\sup(D)$
and (ii) $(A,B) \succ_{L} (C,D)$ iff $\sup(A) > \sup(C)$ or ($\sup(A) = \sup(C)$ and $\sup(B) > \sup(D)$). It is well-known that such a lexicographic ordering has no representation in the real numbers \citep{BEARDON2002}. In other words, there exist no $L:Y \times Y \to \RR$ such that $(A,B) \succsim_{L} (C,D) \iff L(A,B) \geq L(C,D)$. Because orderings and distances are such different objects, it is clear that we need different characterizations for orderings and distances and that is why we handle them separately in Sections~\ref{sec.orderings} and \ref{sec.distances}.

For all $A,B \in Y$, the Hamming distance $H(A,B)$ measures the dissimilarity between $A$ and $B$ and is defined by $H(A,B) = \card{A \triangle B}$. The Hamming distance is a metric and its range  is  $\{0, 1, \ldots, \card{X}  \}$ when $X$ is finite or $\NN \cup \{0\}$ otherwise. Because of this range, the Hamming similarity measure is not $1-H$, but $\card{X}-H$ (when $X$ is finite).

The Jaccard distance $J$ (also named Tanimoto) is defined by 
$$J(A,B) = 
\begin{cases}
1 - \frac{\card{A \cap B}}{\card{A \cup B}}, & \textup {if \ }  A \cup B  \neq \emptyset , \\ 
0,   & \textup {otherwise.} 
\end{cases}
$$
The Jaccard distance  is a metric.
The S\o rensen-Dice distance $S$ (also named Czekanowsky) is defined by 
$$S(A,B) = 
\begin{cases}
1 -  \frac{ 2 \card{A \cap B}}{ \card{A} + \card{B}}, & \textup {if \ }  A \cup B  \neq \emptyset , \\ 
0,   & \textup {otherwise.} 
\end{cases}
$$
The S\o rensen-Dice distance is  a near-metric  because the smallest $\gamma$ for which it satisfies the weak triangle inequality
is $\gamma = 1.5$ \citep{GrageraSuppakitpaisarn2018}.
The Overlap distance (also known as interiority or Szymkiewicz-Simpson) is defined by 
$$O(A,B) = 
\begin{cases}
1 - \frac{\card{A \cap B}}{\min(\card{A} , \card{ B})}, & \textup {if \ }  A  \neq \emptyset \neq B, \\ 
0,   & \textup {if \ }  A  = \emptyset  = B, \\
1,   & \textup {otherwise.}  
\end{cases}
$$
The Overlap distance is not a metric because it fails to satisfy $O(A,B)=0$ iff $A=B$. In particular, $O(A,B)=0$ whenever $A \subseteq B$ or $B \subseteq A$. It is not even a semi-metric or near-metric. To see this, define $A=\{a,b\}$, $B = \{b,c \}$ and $C=\{b\}$. We then have $O(A,B)=1/2, O(A,C)=0$ and $O(C,B)=0$. Hence there is no $\gamma \geq 1$ such that $O(A,B) \leq \gamma \big( O(A,C) + O(C,B)  \big)$. The range of $J,S$ and $O$ is $\QQ \cap [0,1]$ when $X$ is infinite (or a subset thereof otherwise).

\section{Characterization of four dissimilarity orderings}
\label{sec.orderings}

This section presents some characterizations of the orderings induced by the distances introduced in Section~\ref{sec.formal.frame}. We first present three axioms that are satisfied by all orderings discussed in this paper.

\subsection{Common axioms}

The orderings studied in this paper have some very weak and elementary axioms in common. We present them now before characterizing specific orderings.
The first common axiom expresses that the distance or dissimilarity between two sets has no `direction': the distance between $A$ and $B$ is identical to that between $B$ and $A$.
\begin{ax}
\emph{\SYM.} 
\label{A:SYM}
$(A,B) \sim (B,A)$.
\end{ax}
The next  axiom expresses that all elements are equivalent as far as  dissimilarity between sets is concerned. In other words, dissimilarity
 is not affected when we permute the elements.
\begin{ax}
\emph{\NEU.} 
\label{A:NEU} 
$(A,B) \sim (\sigma(A),\sigma(B))$, for all  permutations $\sigma$ of $X$.
\end{ax}
If an ordering satisfies \NEU, all elements of $X$ are kind of clones: in any pair of sets, we can replace any element by another one and we obtain a new pair of sets that is indifferent to the original pair.
If an ordering does not satisfy \NEU, then not all elements are clones, but some of them may be (if they are treated in the same way by $\succsim$). To make this  precise, we formally  define clones. Two elements $a$ and $b$ are clones for $\succsim$ if $(A,B) \sim (\sigma(A), \sigma(B))$ for all $A,B$, where $\sigma$ is the permutation of $X$ such that $\sigma(a)=b, \sigma(b)=a$ and $\sigma(c)=c$ for all $c \in X \setminus \{a,b\}$. Notice that \NEU\ implies that all elements of $X$ are clones.

Although \SYM\ and \NEU\ are weak and basic, they can be questionned. For instance \cite{Tversky1977Sim} finds that dissimilarity judgements made by humans do not obey \SYM.\footnote{To illustrate Tversky's criticism, we 
quote  \citet[][p.~80]{tversky1978}: \textit{We say ``the portrait 
resembles the person'' rather than ``the person resembles the portrait.'' We 
say ``the son resembles the father'' rather than ``the father resembles the 
son.''}.}
 For \NEU, we can  find situations where it is violated. For instance, some  humans in some contexts would probably judge ($\{$Finland$\}, \{$Zimbabwe, Botswana$\} ) \not \sim (\{$Kenya$\}, \{$Malawi, Zambia$\} )$.

The next axiom handles the case of pairs of sets in which both sets are empty.
\begin{ax}
\emph{\TES.} 
\label{A:TES}
For all  $a \in X$,  $(\emptyset, \emptyset) \sim (\{a\},\{a\})$.
\end{ax}
It is difficult to think of a situation where \TES\ would not hold. Moreover, we do not know any dissimilarity ordering in use that violates \TES.

We now turn to the four dissimilarity orderings studied in this paper and we present for each one the additional axioms needed to characterize the ordering.

\subsection{The Hamming ordering}

In order to characterize the Hamming ordering, we introduce three new axioms. The first one says that adding an element to both $A$ and $B$ does not change their dissimilarity.
\begin{ax}
\emph{\IND.} 
\label{A:IND} 
If $c \notin A \cup B$, then $(A \cup \{c\}, B \cup \{c\}) \sim (A,B)$. 
\end{ax}
This axiom is not satisfied by the three other orderings.
The second new axiom examines what happens when we add an element to a set, compared to the empty set. 
\begin{ax}
\emph{\ERH.} 
\label{A:OSEb}
If $c \notin A$, then $(A  \cup \{c\}, \emptyset) \succ (A,\emptyset)$.
\end{ax} 
We will later see other kinds of responsiveness axioms satisfied by other dissimilarities. The suffix `-H' indicates that this axiom is specifically tailored to the Hamming ordering. \ERH\ is not satisfied by the three other orderings.

The third new axiom says that we can move an element from $A\setminus B$ to $B\setminus A$ or vice versa without modifying the dissimilarity.
\begin{ax}
\emph{\TR.} 
\label{A:TR}
If $c \notin A \cup B $, then $(A \cup \{c\}, B) \sim (A, B  \cup \{c\})$. 
\end{ax}
It is  satisfied by $\succsim_{H}$, $\succsim_{J}$ and $\succsim_{S}$, but not by $\succsim_{O}$.
We are now ready to present a characterization of the Hamming ordering.
\begin{theorem}
\label{theo.char.H.ordering}
The ordering $\succsim$ satisfies  \NEU, \TR,  \ERH, and \IND\  iff\, $\succsim$ is the Hamming ordering. 
\end{theorem}
Notice that this result holds irrespective of the cardinality of $X$.
The proof of this result (and of most results) is deferred to Section~\ref{sec.proofs}. The axioms in Theorem~\ref{theo.char.H.ordering} (and in all subsequent theorems) are logically independent.

\subsection{The Jaccard ordering}

For the Jaccard ordering, we  need two new axioms.
The first one shows that the response of the Jaccard ordering to the expansion of set $A$ is  more complex than $\succsim_{H}$'s response.
\begin{ax}
\emph{\ERJ.} 
\label{A:OSE}
If $c \notin A$, then
$$\begin{cases}
(A  \cup \{c\}, B) \succ (A,B), & \textup {if \ }  A \supseteq B \neq \emptyset, \\ 
(A  \cup \{c\}, B) \prec (A,B) ,   & \textup {if \ }  c \in B \supseteq A. 
\end{cases}$$
\end{ax} 
This axiom is satisfied by $\succsim_{H}$, $\succsim_{J}$ and $\succsim_{S}$, but not by $\succsim_{O}$.
The second new axiom essentially says that replicating the elements of $A$ and $B$ has no effect on the dissimilarity.
\begin{ax}
\emph{\RI.} 
\label{A:RI}
For $k \in \NN$ and $i \in \{1, \dots, k\}$, if $f_{i}: A \cup B \to X \setminus (A \cup B)$ are bijections such that 
\begin{itemize}

\item $f_{i}(A \cup B) \cap f_{j}(A \cup B) = \emptyset$ for all $i,j \in \{1, \dots, k\}$ and

\item for each $a \in A \cup B$, $a$ and $f_{i}(a)$ are clones for $\succsim$, 

\end{itemize}
then $(A,B) \sim \left(\bigcup_{i=1}^{k} f_{i}(A) \cup A, \bigcup_{i=1}^{k} f_{i}(B) \cup B \right)$.
\end{ax}

The bijection $f_{i}$ maps each element $a$ of $A \cup B$ on a clone $f_{i}(a)$. The images $f_{1}(a), \ldots, f_{k}(a)$ are $k$ distinct clones of $a$ and $\bigcup_{i=1}^{k} f_{i}(A)$ is a $k$-plication of $A$. The second bullet is not absolutely necessary, but we need it if we want to use logically independent conditions.
\StRI\ is satisfied by $\succsim_{J}$, $\succsim_{S}$ and $\succsim_{O}$, but not by $\succsim_{H}$.
Notice that it blatantly contradicts \IND.

\begin{theorem}
\label{theo.char.J.ordering}
Let $X$ be infinite.
The ordering $\succsim$ satisfies  \NEU, \TR,  \ERJ, \TES, and \StRI\  iff\, $\succsim$ is the Jaccard ordering.
\end{theorem}

\subsection{The S\o rensen-Dice ordering}

We do not need any new axiom for characterizing the S\o rensen-Dice ordering because it is identical to the Jaccard ordering, as expressed in our next result.

\begin{theorem}
\label{theo.char.SD.ordering}
Let $X$ be infinite.
The ordering $\succsim$ satisfies  \NEU, \TR,  \ERJ, \TES, and \StRI\  iff\, $\succsim$ is the S\o rensen-Dice ordering.
\end{theorem}

\bpr
The S\o rensen-Dice ordering satisfies \NEU, \TR, \ERJ, \TES, and \StRI. Hence, by Theorem~\ref{theo.char.J.ordering}, it is  identical to the Jaccard ordering.
\epr

 The identity between the Jaccard and the S\o rensen-Dice ordering was already noticed by \cite{OmhoverDetynieckiRifqiBouchon-Meunier2004}. 
 
\cite{Gerasimou2024} defined a generalized Jaccard distance by
$$J^{\alpha,\beta}(A,B) = 1 - \frac{\card{A \cap B}^{\beta}}{\card{A \cup B}^{\alpha}},$$
where $\alpha$ and $\beta$ are two real numbers such that $1 \geq \alpha \geq \beta > 0$. It is clear that, when $\alpha=\beta$, 
$J^{\alpha,\beta}$ is a strictly increasing transformation of $J$. Hence the ordering induced by $J^{\alpha,\alpha}$ is identical to $\succsim_{J}$ and the five axioms of Theorem~\ref{theo.char.J.ordering} or \ref{theo.char.SD.ordering} also characterize the ordering induced by $J^{\alpha,\alpha}$. When $\alpha \ne \beta$, $J^{\alpha,\beta}$ violates \StRI.
 
 Notice that Theorem~\ref{theo.char.SD.ordering} does not imply that the Jaccard distance (or $J^{\alpha,\alpha}$) is identical to the S\o rensen-Dice distance.

\subsection{The Overlap ordering}

For the Overlap ordering, we need three new axioms. The first two examine what happens when we add an element to a set. They show that  $\succsim_{O}$'s response to the expansion of set $A$ is even more complex than $\succsim_{J}$'s response.

\begin{ax}
\emph{\ERO.} 
\label{A:ER} 
If $c \notin A \cup B$, then 
$$
\begin{cases}
 (A \cup \{c\},B) \succ (A,B), &\textrm{if } \card{A} < \card{B} \textrm{ and } A \cap B \neq \emptyset, \\
 (A,B) \succ (A \cup \{c\}, B \cup \{c\}), &\textrm{if }  A \nsubseteq B  \textrm{ and } B \nsubseteq A.
\end{cases}
$$
\end{ax}
It is satisfied by $\succsim_{J}$, $\succsim_{S}$ and $\succsim_{O}$, but not by $\succsim_{H}$.

\begin{ax}
\emph{\EI.} 
\label{A:EI} 
If $c \notin A \cup B$, $A \cup B \neq \emptyset$ and $\card{A} \geq  \card{B}$, then $(A \cup \{c\},B) \sim (A,B)$.
\end{ax}
It is satisfied by $\succsim_{O}$, but not by the other three orderings.
When adding an element  to only one of the sets (say $A$), \EI\ imposes that the dissimilarity does not increase if  $\card{A} \geq \card{B}$. Consequently, if $A \supseteq B$, then enlarging $A$ has no effect on the dissimilarity. In particular, all pairs $A,B$ such that $A \supseteq B$ have the lowest position in the ordering $\succsim_{O}$ because $O(A,B)=0$.

The third new axiom  handles the cases of pairs of sets in which one  set is empty.
\begin{ax}
\emph{\OES.} 
\label{A:OES} 
For all distinct $a,b \in X$, $(\{a \}, \emptyset) \sim (\{a \}, \{ b \})$.
\end{ax}
Although we did not introduce \OES\ earlier, it is also satisfied by $\succsim_{J}$ and $\succsim_{S}$, but not by $\succsim_{H}$.
This axiom is very weak, but not completely inocuous. For instance, it is likely that, in some contexts, some individuals will judge ($\{$Finland$\}, \emptyset ) \not \sim (\{$Finland$\}, \{$Zimbabwe$\} )$.

\begin{theorem}
\label{theo.char.O.ordering}
Let $X$ be infinite.
The ordering $\succsim$ satisfies \NEU, \SYM,  \StRI, \OES, \TES, \EI\  and \ERO\  iff\, $\succsim$ is the Overlap ordering.
\end{theorem}

\section{Characterization of four distances}
\label{sec.distances}

In Section~\ref{sec.orderings}, we have defined some axioms for dissimilarity orderings. We say that the distance $I$ satisfies an axiom defined for dissimilarity orderings if the induced dissimilarity ordering $\succsim_{I}$ satisfies the axiom. For instance, the ordering induced by $H$  satisfies \SYM\ and we therefore say that $H$ itself satisfies \SYM.

\subsection{Common axioms}

The four distances considered in this paper satisfy two additional common axioms. The first one  seems very uncontroversial.
\begin{ax}
\emph{\LB.} 
\label{A:LB}
For all $a \in X$,  $I(\{a\}, \{a\})=0$. 
\end{ax}
It is similar to, but weaker than the third condition in the definition of distances (Section~\ref{sec.formal.frame}).
The second common axiom is often imposed on distances, probably for practical reasons.
\begin{ax}
\emph{\Unit.} 
\label{A:Unit}
For all  $a \in X$, we have $I(\{a\}, \emptyset) =1$. 
\end{ax}
Notice that the choice of the number 1 as unit is  arbitrary and not compelling.

\subsection{The Hamming distance}

Any strictly increasing transformation of the Hamming distance $H$ induces the same ordering $\succsim_{H}$. The four axioms of Theorem~\ref{theo.char.H.ordering} are therefore too weak to characterize $H$. We need  one additional axiom.

\begin{ax}
\emph{\ADD.} 
\label{A:ADD} 
$A \cup B \neq \emptyset \Rightarrow I(A,B) =  I(A, A \cup B) + I(A \cup B, B)$. 
\end{ax}
This axiom imposes  that the distance between $A$ and $B$ can be additively decomposed in two parts: on the one hand the distance between $A$ and $A \cup B$, and on the other hand between $A \cup B$ and $B$.\footnote{The Hamming distance also satisfies another additivity condition: $A \cup B \neq \emptyset \Rightarrow I(A,B) =  I(A, A \cap B) + I(A \cap B, B)$.\label{f.add}}

\begin{theorem}
\label{theo.char.H.dist}
The mapping $I:Y \times Y \to \RR$ satisfies  \NEU, \TR,  \ERH, \IND\ and \ADD\ iff $I$ is the Hamming distance (up to a scale factor), that is $I = \alpha H$ for some positive  $\alpha \in \RR$.
\end{theorem}
Most people use $H$ and not $\alpha H$, but this is just a matter of convenience. It is as arbitrary as measuring distances in inches instead of meters. For this reason, we consider Theorem~\ref{theo.char.H.dist} as a complete characterization of the Hamming distance.
If we nevertheless wish to single out the Hamming distance without the scale factor, we can impose \Unit.

\begin{corollary}
\label{theo.char.H.dist.UB}
The mapping $I:Y \times Y \to \RR$ satisfies  \NEU, \TR,  \ERH, \IND, \ADD\ and \Unit\ iff $I$ is the Hamming distance.
\end{corollary}

So far, we have not used the triangle inequality.\footnote{\cite{Tversky1982similarity_traingleLaw}  criticized the triangle inequality which 
may not be applicable in many psychological studies.}
 It is not strong enough to replace \ADD, but it does the job if we combine it with a super-additivity axiom.

\begin{ax}
\emph{\SA.} 
\label{A:SA} 
$I(A,B) \geq   I(A, A \cup B) + I(A \cup B, B)$. 
\end{ax}

\begin{theorem}
\label{theo.char.H.dist.tr}
The mapping $I:Y \times Y \to \RR$ satisfies  \NEU, \TR,  \ERH,  \IND, \SA\ and the triangle inequality iff $I$ is the Hamming distance (up to a scale factor), that is $I = \alpha H$ for some positive  $\alpha \in \RR$.
\end{theorem}

We will see in Section~\ref{subsec.Sorensen} and~\ref{subsec.Overlap} that the S\o rensen-Dice and Overlap distances do not satisfy \ADD. In order to characterize them, we will need  constant sensitivity axioms. We therefore find it useful to present here a second characterization of $H$ using  one of these constant sensitivity axioms. This  characterization of $H$ will be amenable to comparisons with the characterizations of $S$ and $O$.
\begin{ax}
\emph{\TSES.}  
\label{A:TSES}
If $a$ and $b \in B \setminus A$ are clones for $\succsim_{I}$ and $\card{A} = \card{B}$, then $I(A,B)- I(A \cup \{a\},B ) = I(A \cup \{a\},B ) - I(A \cup \{a,b\},B )$.
\end{ax}

The suffix `-O' indicates that this axiom is specifically tailored to the Overlap distance, although it is also satisfied by the Hamming and Jaccard distances.\footnote{The Hamming and Jaccard distances satisfy \StTSES\ even if we drop the restriction $\card{A} = \card{B}$. The Overlap distance does not.}

\begin{theorem}
\label{theo.char.H.dist2}
The mapping $I:Y \times Y \to \RR$ satisfies  \NEU, \TR,  \ERH, \IND, \LB\ and \TSES\ iff $I$ is the Hamming distance (up to a scale factor), that is $I = \alpha H$ for some positive  $\alpha \in \RR$. \end{theorem}

Theorem~\ref{theo.char.H.dist2} can be proved with a variant of \TSES\ omitting the premise that $a$ and $b$ be clones. Yet, this premise is needed if we want to use logically independent conditions.

\subsection{The Jaccard distance}

The Jaccard distance, just like the Hamming distance, satisfies \ADD.\footnote{The Jaccard distance also satisfies the alternative additivity condition presented in footnote~\ref{f.add}.}
 This leads us to Theorem~\ref{theo.char.J.dist}, which is the Jaccard analogue to Theorem~\ref{theo.char.H.dist} and therefore makes the comparison of $H$ and $J$ very easy.

\begin{theorem}
\label{theo.char.J.dist}
Let $X$ be infinite.
The mapping $I:Y \times Y \to \RR$ satisfies  \NEU, \TR,  \ERJ, \TES, \StRI\ and \ADD\ iff $I$ is the Jaccard distance (up to a scale factor), that is $I = \alpha J$ for some positive  $\alpha \in \RR$.
\end{theorem}
As for the Hamming distance, we can impose  \Unit\ to obtain $I=J$ and we can replace \ADD\ by the conjunction of \SA\ and the triangle inequality. The formal statement of these results is omitted.

As for the Hamming distance, we present an alternative characterization using \StTSES\ instead of \ADD.

\begin{theorem}
\label{theo.char.J.dist2}
Let $X$ be infinite.
The mapping $I:Y \times Y \to \RR$ satisfies  \NEU, \TR,  \ERJ, \TES, \StRI, \LB\ and \StTSES\ iff $I$ is the Jaccard distance (up to a scale factor), that is $I = \alpha J$ for some positive  $\alpha \in \RR$. 
\end{theorem}

\subsection{The S\o rensen-Dice distance}
\label{subsec.Sorensen}

\ADD\ is violated by the S\o rensen-Dice distance, but \ADD\ is just one of the many additivity conditions that we could write. For instance, $I(A,B)=I(A, A \cap B) + I(A \cap B, B)$ or $I(A,B)=I(A \setminus B, A \cup B) + I(A \cup B, B \setminus A)$, etc. We propose the following definition of a general additivity condition.
\begin{ax}
\emph{\GADD.}  
\label{A:GADD}
$I$ satisfies \GADD\ if there are four mappings $\kappa, \lambda, \mu, \nu: Y^{2} \to Y$ such that,  for all $A,B \in Y$,
\begin{itemize}  
\item $\kappa(A,B)$ (resp. $\lambda, \mu, \nu$) can be written in terms of $A,B, \cup, \cap, \setminus$;
\item $I(A,B)= I\big(\kappa(A,B), \lambda(A,B) \big) + I\big(\mu(A,B), \nu(A,B) \big)$;  
\item $\min\Big(I\big(\kappa(A,B), \lambda(A,B) \big),I\big(\mu(A,B), \nu(A,B) \big)\Big)>0$ for some $A,B$.
\end{itemize}
\end{ax}
The third part of the definition excludes trivial additivity conditions like $I(A,B)=I(A,B)+I(A,A)$. 
 \ADD\ is an instance of \GADD, in which $\kappa(A,B)=A$, $\lambda(A,B)=A \cup B = \mu(A,B)$ and $\nu(A,B)=B$. 
Our next result shows that the S\o rensen-Dice distance does not satisfy any kind of additivity.
\begin{theorem}
\label{theo.S.no.add}
The S\o rensen-Dice distance violates \GADD.
\end{theorem}
We will therefore provide only one characterization of the S\o rensen distance, using a new constant sensitivity axiom.
\begin{ax}
\emph{\TSESS.}  
\label{A:TSESS} 
If $a$ and $c$ (resp.\ $b$ and $d$) are clones for $\succsim$ such that
$a, b \notin A \cup B$ and  $c,d \in A \cap B$, 
then $I(A ,B) -  I(A \cup \{a\},B\setminus \{c\}) =
 I(A \cup \{a\},B\setminus \{c\}) - I(A \cup \{a,b\},B \setminus \{c,d\})$.
\end{ax}

The suffix `-S' indicates that this axiom is specifically tailored to the S\o rensen distance. It is not satisfied by $H,J$ nor $O$. As for \TSES, the condition that $a$ and $c$ (resp.\ $b$ and $d$) be clones is not absolutely necessary but is needed if we want to use logically independent conditions.

\begin{theorem}
\label{theo.char.S.dist}
Let $X$ be infinite.
The mapping $I:Y \times Y \to \RR$ satisfies  \NEU, \TR,  \ERJ, \TES, \StRI, \LB\ and \StTSESS\ iff $I$ is the S\o rensen-Dice distance (up to a scale factor), that is $I = \alpha S$ for some positive $\alpha \in \RR$.
\end{theorem}

\subsection{The Overlap distance}
\label{subsec.Overlap}

The Overlap distance also violates \GADD\ (and, hence, \ADD).
\begin{theorem}
\label{theo.O.no.add}
The Overlap distance violates \GADD.
\end{theorem}
Yet, it satisfies \StTSES\ and this leads us to our next result.

\begin{theorem}
\label{theo.char.O.dist}
Let $X$ be infinite.
The mapping $I:Y \times Y \to \RR$ satisfies  \NEU, \SYM, \StRI, \OES, \TES, \EI, \ERO, \LB\ and \StTSES\ iff $I$ is the Overlap distance (up to a scale factor), that is $I = \alpha O$ for some positive $\alpha \in \RR$.
\end{theorem}
%

\section{Discussion}
\label{sec.discussion}

\subsection{Summary}

Table~\ref{table} provides a summary of the results. Since many of the axioms satisfied by a distance are also satisfied by other distances, we  provide hereafter a classification of the four distances according to three distinctive properties.

The Jaccard, S\o rensen and Overlap distances all satisfy \StRI.  On the contrary, the Hamming distance is rather an extensive concept and that is why it satisfies \IND. 
The Hamming, Jaccard and S\o rensen distances all satisfy \ERJ. The response of $O$ to the expansion of a set is more complex.
The Hamming and Jaccard distances are the only ones satisfying \ADD, which seems to be a more appealing axiom than any of the constant sensitivity axioms.

\begin{figure}[htbp]
    \centering 
    \scalebox{0.5}{
    \includegraphics{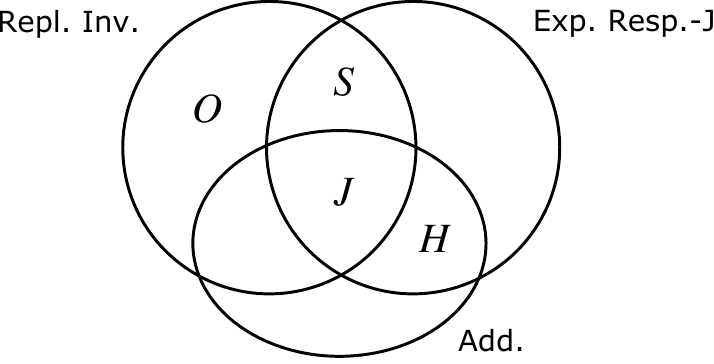}
    }
    \caption{Summary of the results.}
    \label{fig.distances}
\end{figure}

These facts are visually represented in Fig.~\ref{fig.distances} and can be formally expressed as our final result.
\begin{theorem}
\label{theo.summary}
Among the family $\{H,J,S,O \}$, 
\begin{itemize}
\item $H$ is the only distance satisfying \ADD, and \ERJsh, but not \StRI.
\item $J$ is the only distance satisfying \ADD, \StRI\ and \ERJsh.
\item $S$ is the only distance satisfying  \StRI\ and \ERJsh, but not \ADD.
\item $O$ is the only distance satisfying  \StRI, but not \ADD\ and \ERJsh.
\end{itemize}
\end{theorem}

\begin{table}
\begin{tabular}{lllllllllllllllll}
\hline
 & \rb{\SYM} & \rb{\TR} & \rb{\NEU} & \rb{\TES} & \rb{\OES} & \rb{\IND} & \rb{\StRIsh} & \rb{\ERHsh} & \rb{\ERJsh} & \rb{\EROsh} & \rb{\EIsh} & \rb{\LB} & \rb{\Unit} & \rb{\ADD} & \rb{\StTSESsh} & \rb{\StTSESSsh}  \\
\hline
$H$ & \xm	 & \cm & \cm & \xm &   & \cm & & \cm & \xm & & & \xm & \xm & \cm & \xm & \xm     \\ 
$J$ & \xm	 & \cm & \cm & \cm & \xm &  & \cm &  & \cm & \xm & & \xm & \xm & \cm & \xm &     \\ 
$S$ & \xm	 & \cm& \cm & \cm & \xm  & & \cm &  & \cm & \xm & & \cm & \xm & & & \cm     \\ 
$O$ & \cm &  & \cm & \cm & \cm & & \cm &  &  & \cm & \cm & \cm & \xm & & \cm &     \\ 
\hline		
\end{tabular}
\caption{Summary of the results. The axioms characterizing $H$ (resp.~$J, S, O$) in Theorem~\ref{theo.char.H.dist} (resp.\ \ref{theo.char.J.dist}, \ref{theo.char.S.dist}, \ref{theo.char.O.dist}) are marked by \cm\ in the corresponding row. The axioms satisfied by a distance are marked by \xm or \cm. An empty cell indicates that the axiom is not satisfied by the corresponding distance.}
\label{table}
\end{table}

\subsection{Comparison with the axioms in \cite{Gerasimou2024}}

The characterization of $J$ by \cite{Gerasimou2024} uses three axioms: 
\begin{description}
\item [A1] $I(A,B)=I(B,A)$,

\item [A2] $I(A,B)= 0$ iff $A=B$ and

\item [A3] $x \in B \setminus A \implies I(A,B) - I(A \cup \{x\}, B) = 1/ \card{A \cup B}$.
\end{description}

Our result that most closely resembles Gerasimou's characterization is Theorem~\ref{theo.char.J.dist2}. It involves   \NEU, \TR,  \ERJ, \TES, \StRI, \LB\ and \StTSES. Notice that we need \TES\ simply because we characterize $J$ on the full domain, while Gerasimou excludes the case of two empty sets.
We now compare Gerasimou's conditions and ours.

Gerasimou's A1 is quite a weak condition, but, in Theorem~\ref{theo.char.J.dist2}, we have decomposed A1 into  two even weaker conditions (\NEU\ and \TR) that, together, imply A1.

Gerasimou's A2 is also rather weak. It implies our \LB\ and, on the full domain, \TES, while the conjunction of these two conditions does not even imply A2.

If we apply twice Gerasimou's A3, with $a,b \in B \setminus A$, we obtain $I(A,B) - I(A \cup \{a\}, B) = I(A \cup \{a\}, B) - I(A \cup \{a,b\}, B) = 1/ \card{A \cup B}$, which looks very similar to our \TSES. But A3 is stronger because it not only imposes a constant sensitivity, but also  that the sensitivity be $1/ \card{A \cup B}$.\footnote{\TSES\ is  a strengthening of A3' in \cite{Gerasimou2024}.}

In conclusion, we use weaker axioms than Gerasimou and that is why we need more axioms to characterize $J$. The benefit of using more and weaker axioms is that they are more elementary: each axiom is easier to interpret; the normative or descriptive content of each axiom is simpler. This is particularly salient when we compare the very strong  A3 with our axioms.

\subsection{Cardinality of $X$}

Another difference with Gerasimou's paper resides in the cardinality constraints imposed on
 the universal set $X$. Throughout his paper, Gerasimou imposes that $X$ be finite. In our paper, all results pertaining to the Hamming distance  ordering are proven without any cardinality constraint: $X$ can be finite, countably infinite or uncountable. In our other results, the proof technique requires $X$ to be infinite (countable or not). This makes our results about the Jaccard distance somehow complementary to Gerasimou's result.
 
 Notice that, although the universal set $X$ is infinite in some of our results, the distance is  only defined for finite subsets of $X$. This perfectly reflects what is done in some applications. For instance, in scientometrics \citep{Baccinietal2024}, dissimilarities are computed between papers. More precisely, the dissimilarity between papers $p$ and $q$ is the distance between the set $A$ of papers cited by $p$ and the set $B$ of papers cited by $q$.
The universal set $X$ is therefore the set of all papers in the database. One can argue that, although the size of the database constantly increases, every computation will use a finite database. This is true, but we do not want that the computations made today  or tomorrow (with a larger database) bear no relation to each other. So, when characterizing a distance, we have to anticipate that $X$ will keep growing and, since we cannot put any bound on $\card{X}$, we have to consider $X$ as infinite.

In chemical informatics \cite{Maggiora2014}, $X$ is the set of features that molecules possess or not and distances are then used to compute the dissimilarity between molecules. Since the number of possible molecules is potentially unbounded, we need infinitely many features to distinguish them all. Of course, when computing dissimilarities between all molecules (sets of features) in a given database, only finite sets are involved, but the underlying universal set $X$ of features is infinite.

\section{Proofs}
\label{sec.proofs}

This section contains the proofs of all results presented in the paper. 

\subsection{Preliminary lemmas}

Given a pair of sets $(A,B)$, we define its type as a vector of three non-negative integers: 
$$t^{AB}=
\begin{cases}
\big(\card{A \setminus B}, \card{A \cap B}, \card{B \setminus A } \big), & \textup {if \ }  \card{A} \geq \card{B}, \\ 
\big(\card{B \setminus A}, \card{A \cap B}, \card{A \setminus B } \big), & \textup {if \ }  \card{A} < \card{B }. 
\end{cases}$$
For instance, if $A= \{a,b,c\}$ and $B=\{c,d\}$, then $t^{AB}=(2,1,1)$ and $t^{BA}=(2,1,1)$. 

\begin{lemma}
\label{lemma.type}
If the ordering $\succsim$ satisfies \NEU\  and  \SYM, then all pairs with the same type are indifferent.
\end{lemma}

\bpr
Consider four sets $A,B,C,D$ such that $t^{AB}=t^{CD}$. By definition of the type, $\card{A}=\card{C}$ or $\card{A}=\card{D}$. Thanks to \SYM, we can assume $\card{A}=\card{C}$ without loss of generality. Hence $\card{B}=\card{D}$ and there is a permutation $\sigma$ of $X$ such that $\sigma(A)=C$ and $\sigma(B)=D$. Neutrality  imposes $(A,B) \sim (C,D)$
\epr

A consequence of this lemma is that an ordering $\succsim$ is completely defined if we order all types. This is what we will do in most proofs. We will often abuse the notation and write $(i,j,k) \succsim (i',j',k')$ when we mean that $(A,B) \succsim (C,D)$ for all sets $A,B,C,D$ such that $t^{AB}= (i,j,k)$ and $t^{CD}=(i',j',k')$. 

Notice that the set of all types is a subset of $(\NN \cup \{0\})^{3}$. This set is countable, even when $X$ is uncountable. Hence, by Lemma~\ref{lemma.type}, each dissimilarity ordering satisfying \NEU\  and  \SYM\ admits a numerical representation $I$ as in \eqref{num.repr}.

\begin{lemma}
\label{lemma.SYM}
If the ordering $\succsim$ satisfies \NEU\ and \TR, then it satisfies \SYM.
\end{lemma}

\bpr
Let $A,B \in Y$ with $\card{A \setminus B} = i, \card{A \cap  B} = j$ and $\card{B \setminus A} = k$.  If $i=k$, then \NEU\ clearly implies $(A,B) \sim (B,A)$.
Hence, suppose without loss of generality $i > k$. Using \TR\ $(i-k)$ times, we can `move' $(i-k)$ elements from $A \setminus B$ to $B \setminus A$. We obtain two sets $A',B'$ such that $\card{A' \setminus B'} = k, \card{A' \cap  B'} = j$ and $\card{B' \setminus A'} = k$. By \TR, $(A,B) \sim (A',B')$ and, by \NEU, $(A',B') \sim (B,A)$. By transitivity, $(A,B) \sim (B,A)$.
\epr

\subsection{Hamming}

\noindent {\bf Proof of Theorem~\ref{theo.char.H.ordering}.} 
The proof of necessity is simple and omitted. Let us show the sufficiency. Consider any $A,B, C,D \in Y$ with types $x=t^{AB}$ and $y = t^{CD}$ and suppose without loss of generality $H(A,B) \geq H(C,D)$.  We must show that $(A,B) \succsim (C,D)$.  By \IND, \NEU\ and \TR, we clearly have $(x_{1},x_{2},x_{3}) \sim (x_{1},0,x_{3}) \sim (x_{1}+x_{3},0,0)$. Similarly, $(y_{1},y_{2},y_{3}) \sim (y_{1},0,y_{3}) \sim (y_{1}+y_{3},0,0)$.
\begin{itemize}

\item If $H(A,B) = H(C,D)$, then $x_{1}+x_{3} = y_{1}+y_{3}$ and transitivity implies $(x_{1},x_{2},x_{3}) \sim (x_{1},0,x_{3}) \sim (x_{1}+x_{3},0,0) = (y_{1}+y_{3},0,0) \sim (y_{1},0,y_{3}) \sim (y_{1},y_{2},y_{3})$.

\item If $H(A,B) > H(C,D)$, then $x_{1}+x_{3} > y_{1}+y_{3}$. Then transitivity and \ERH\ imply $(x_{1},x_{2},x_{3}) \sim (x_{1},0,x_{3}) \sim (x_{1}+x_{3},0,0) \succ (y_{1}+y_{3},0,0) \sim (y_{1},0,y_{3}) \sim (y_{1},y_{2},y_{3})$. 

\end{itemize}
The logical independence of the axioms of Theorem~\ref{theo.char.H.ordering} will be established in the proof of Theorem~\ref{theo.char.H.dist} about the Hamming distance.
\epr

\noindent {\bf Proof of Theorem~\ref{theo.char.H.dist}.} 
It is easy to check that the Hamming distance satisfies \ADD. We turn to the sufficiency part. 
Thanks to Theorem~\ref{theo.char.H.ordering}, we know that $I$ is a numerical representation of $\succsim_{H}$. There exists therefore $\phi:  \{0, \ldots, \card{X}\} \to \RR$, strictly increasing, such that $I = \phi(H)$. Thanks to Lemma~\ref{lemma.type}, \ADD\ can be written as $I(i,j,k) = I(i,j+k,0) + I(k,i+j,0)$. So, 
$\phi(i+k) = \phi(i) + \phi(k)$, for all $i,k \in \{0, \ldots, \card{X}\}$ with $i+k \leq \card{X}$. Setting $i=k=0$, $\phi(0)=0$ obtains. Setting $k=1$ and $i= 1,2, 3, \ldots$, we have
\begin{align*}
\phi(1+1)  & = 2 \phi(1) , \\
\phi(2+1)  & =  \phi(2) + \phi(1) = 3 \phi(1), \\
\phi(3+1)  & =   \phi(3)+ \phi(1) = 4 \phi(1) , \\
\ldots 
\end{align*}
Hence, $\phi(i) = i \phi(1)$ for all $i \in \{0, \ldots, \card{X}\}$.
Put differently, $I=\phi(1) H$.

Let us now prove the logical independence of the axioms. For each axiom, we present a distance satisfying all axioms but one.

\noindent \IND: $J$.

\noindent  \TR: 
$I_{\ref{ex.H.no.TR}}(A,B)= 2 \card{A \setminus B} + \card{B \setminus A}$. 
\refstepcounter{myindex} 
\label{ex.H.no.TR}

\noindent \ERH: $-H$ or 
\refstepcounter{myindex} 
\label{ex.H.d.no.OSE}
  $I_{\ref{ex.H.d.no.OSE}}(A,B)=0$ for all $A,B \in Y$.

\noindent \NEU:
$I_{\ref{ex.H.no.NEU}}(A,B)=\sum_{a \in A \triangle B}w(a)$ with $w:X \to \NN$  a non-constant mapping. \refstepcounter{myindex}  \label{ex.H.no.NEU} 

\noindent \ADD:
$H-1$. 
\epr

\noindent {\bf Proof of Theorem~\ref{theo.char.H.dist.tr}.} 
The triangle inequality implies $I(A,B) \leq  I(A, A \cup B) + I(A \cup B, B)$. This and \SA\ yields \ADD. We can therefore apply Theorem~\ref{theo.char.H.dist}.

In order to prove the logical independence of the axioms, we can use the same examples as in Theorem~\ref{theo.char.H.dist} for \NEU, \TR, \ERH\ and \IND. For the remaining two axioms, we need  new examples. 

\noindent \SA: $H^{1/2}$.

\noindent Triangle inequality: $H^{2}$.
\epr

\noindent {\bf Proof of Theorem~\ref{theo.char.H.dist2}.} 
The necessity part is easy and is thus omitted.
 Thanks  to Lemma~\ref{lemma.type},   \TSES\ can  be rewritten as
 $I(i,j,i)-I(i,j+1,i-1) = I(i,j+1,i-1)-I(i,j+2,i-2)$.
 Thanks to Theorem~\ref{theo.char.H.ordering}, we know that $I$ is a numerical representation of $\succsim_{H}$. There exists therefore $\phi: \{0, \ldots, \card{X}\} \to \RR$, strictly increasing, such that $I = \phi(H)$. Hence, 
$$\phi( i+i )- \phi(i+i-1) = \phi(i+i-1) - \phi(i+i-2),$$
for all $i \in \{0, \ldots, \card{X}\}$ with $1 \leq i \leq \card{X}/2$. With $z=2i$, we have  $\phi( z ) = 2 \phi(z-1) - \phi(z-2)$. 
 Writing this condition for $z = 2, 3,  \ldots$, we find 
\begin{align*}
\phi(2)  & = 2 \phi(1) - \phi(0) = 2 \big(\phi(1) - \phi(0) \big) + \phi(0), \\
\phi(3)  & = 2 \phi(2) - \phi(1) = 3 \big(\phi(1) - \phi(0) \big) + \phi(0), \\
\phi(4)  & =  2 \phi(3) - \phi(2) = 4 \big(\phi(1) - \phi(0) \big) + \phi(0),  \\
\ldots 
\end{align*}
Hence, for all $z \in \{0, \ldots, \card{X}\}$, $\phi(z)= z \big(\phi(1)-\phi(0) \big)+\phi(0)$.  By \LB, 
$\phi(0)=0$. This implies $\phi(z)= z \phi(1)$ and  $I=\phi(1)H$.

We now turn to the proof of the logical independence of the conditions.

\noindent \NEU:
$I_{\ref{ex.H.no.NEU}}$. 

\noindent \TR: $I_{\ref{ex.H.no.TR}}$.

\noindent \ERH:  $I_{\ref{ex.H.d.no.OSE}}$ or $-H$.

\noindent \IND:
$J$.

\noindent \LB:
$H+1$

\noindent \TSES:
$H^{2}$.
\epr

\subsection{Jaccard}

We first  prove a lemma that will also be useful for the characterizations of the S\o rensen-Dice and Overlap orderings.
\begin{lemma}
	\label{lemma.RI}
	Let $X$ be infinite. If the ordering $\succsim$ satisfies \NEU\ and \RI, then, for any $\theta \in \NN$, we have $(i,j,k)\sim (\theta i,\theta j,\theta k)$.
\end{lemma}

\bpr
Consider any $A,B \in Y$. By \NEU,  for any bijection $f_{i}: A \cup B \to X \setminus (A \cup B)$, we have  $(\{a, f_{i}(a)\} , \{a\}) \sim (\{a, f_{i}(a)\} , \{f_{i}(a)\})$, for each $a \in A \cup B$.  Since $X$ is infinite, we can then  apply \RI\ without any restriction, and the rest of the proof easily follows.
\epr

\noindent {\bf Proof of Theorem~\ref{theo.char.J.ordering}.} 
Showing that the Jaccard ordering satisfies all the axioms is easy  and is  omitted. 

Now suppose that  the ordering $\succsim$ satisfies all axioms of Theorem~\ref{theo.char.J.ordering}, then we will show that $\succsim$ is the Jaccard ordering.
Consider any $A,B,C,D \in Y$ with types $x=t^{AB}$ and $y = t^{CD}$ and suppose without loss of generality $J(A,B) \geq J(C,D)$.  We must show that $(A,B) \succsim (C,D)$. 

Let us first consider the cases where $(A,B)$ or $(C,D)$ is equal to $(\emptyset,\emptyset)$. 
\begin{itemize}
\item If $(C,D)=(\emptyset,\emptyset)=(A,B)$, then obviously $(A,B) \sim (C,D)$. 

\item  If $(C,D)=(\emptyset,\emptyset)\neq (A,B)$, then $(C,D) \sim (0,0,0) \sim (0,1,0)$, by Lemma~\ref{lemma.type} and \TES. By \RI, $(0,1,0) \sim (0,x_{2},0)$. By \ERJ, $(0,x_{2},0) \precsim (x_{1}+x_{3},x_{2},0)$ (the comparison is not strict because $x_{1}+x_{3}$ can be zero). By \TR, 
$(x_{1}+x_{3},x_{2},0) \sim (x_{1},x_{2},x_{3})$. By transitivity, $(C,D) \precsim (A,B)$.

\item If $(A,B)=(\emptyset,\emptyset)\neq (C,D)$, then $J(A,B) \geq J(C,D)$ implies $J(C,D)=0$ and, hence, $C=D$. Then, by \RI\ and \NEU,
$(C,D) \sim (\{a \},\{ a \})$, with $a$ as in the statement of \TES. By \TES, $(\{ a \},\{ a \}) \sim (\emptyset,\emptyset)$. By transitivity, $(C,D) \sim (A,B)$.

\end{itemize}

The rest of the proof assumes $(A,B) \neq (\emptyset,\emptyset) \neq (C,D)$.
By \TR, $x \sim (x_{1}+x_{3},x_{2},0)$.
By \RI, if $y_{2} > 0$, then we have $x \sim (x_{1}+x_{3},x_{2},0) \sim (y_{2}(x_{1}+x_{3}),y_{2}x_{2},0)$ and, similarly, if $x_{2} > 0$, $y \sim (y_{1}+y_{3},y_{2},0)  \sim (x_{2}(y_{1}+y_{3}),x_{2}y_{2},0)$. Because $J(A,B) \geq J(C,D)$, if $x_{1}+x_{3} \neq 0 \neq y_{1}+y_{3}$, then
$$\frac{x_{2}}{x_{1}+x_{3}} \leq \frac{y_{2}}{y_{1}+y_{3}} $$
and $x_{2} (y_{1}+y_{3}) \leq y_{2} (x_{1}+x_{3}).$

\begin{itemize}

\item If $J(A,B) = J(C,D)=0$, then $A=B$ and $C=D$. Then Lemma~\ref{lemma.type} and \RI\ imply $(A,B) \sim (0,x_{2},0) \sim (0,y_{2},0) \sim (C,D)$.

\item If $1 > J(A,B) = J(C,D) > 0$, then $x_{2} (y_{1}+y_{3}) = y_{2} (x_{1}+x_{3})$ and, hence,
$x \sim (x_{1}+x_{3},x_{2},0) \sim (y_{2}(x_{1}+x_{3}),y_{2}x_{2},0) = (x_{2}(y_{1}+y_{3}),x_{2}y_{2},0) \sim (y_{1}+y_{3},y_{2},0) \sim y$. Notice that the type $(y_{2}(x_{1}+x_{3}),y_{2}x_{2},0)$ corresponds to an existing pair of sets because we have assumed $X$ is infinite.

\item If $J(A,B) = J(C,D)=1$, then $x_{2}  = y_{2} =0$ and \RI\ implies
$x \sim (x_{1}+x_{3},0,0) \sim (1,0,0) \sim (y_{1}+y_{3},0,0) \sim y$.

\item If $1 > J(A,B) > J(C,D)=0$, then $C=D$. By \TR\ and  successive applications of \ERJ, $(A,B) \sim (x_{1},x_{2},x_{3}) \sim (x_{1}+x_{3},x_{2},0) \succ (0, x_{2}, 0)$. By \RI, $(0, x_{2}, 0) \sim (0, y_{2}, 0) \sim (C,D)$. By transitivity, $(A,B) \succ (C,D)$.

\item If $1 > J(A,B) > J(C,D) >0$, then $x_{2} (y_{1}+y_{3}) < y_{2} (x_{1}+x_{3})$. So, by successive applications of \ERJ, $(y_{2}(x_{1}+x_{3}),y_{2}x_{2},0) \succ (x_{2}(y_{1}+y_{3}),x_{2}y_{2},0)$. By transitivity and \RI, $x \sim (y_{2}(x_{1}+x_{3}),y_{2}x_{2},0) \succ (x_{2}(y_{1}+y_{3}),x_{2}y_{2},0) \sim y$.

\item  If $1 = J(A,B) > J(C,D)=0$, then $C=D$ and $x_{2}=0$. By \TR\ and \RI, $(A,B) \sim (x_{1},0,x_{3}) \sim (x_{1} + x_{3},0, 0) \sim (y_{2},0,0)$. By \ERJ, $(y_{2},0,0) \succ (0, y_{2},0) \sim (C,D)$. By transitivity, $(A,B) \succ (C,D)$.

\item If $1 = J(A,B) > J(C,D) >0$, then $y_{2} > x_{2} =0$. By \RI, $x \sim (x_{1}+x_{3},0,0) \sim (y_{1}+y_{3}+y_{2},0,0)$. By \ERJ, $(y_{1}+y_{3}+y_{2},0,0) \succ (y_{1}+y_{3},y_{2},0) \sim y$.
By transitivity, $x \succ  y$.

\end{itemize}
The logical independence of the axioms of Theorem~\ref{theo.char.J.ordering} will be established  in the proof of Theorem~\ref{theo.char.J.dist} about the Jaccard distance.
\epr

\noindent {\bf Proof of Theorem~\ref{theo.char.J.dist}.} 
It is easy to check that the Jaccard distance satisfies \ADD. We turn to the sufficiency part. 
Thanks to Theorem~\ref{theo.char.J.ordering}, we know that $I$ is a numerical representation of $\succsim_{J}$. There exists therefore $\phi: \RR \to \RR$, strictly increasing, such that $I = \phi(J)$. The rest of the proof of sufficiency follows that of Theorem~\ref{theo.char.H.dist}.

We now prove the logical independence of the axioms.

\noindent \NEU:
Let $w:X \to \NN$ be a a non-constant mapping and 
$$I_{\ref{ex.J.no.NEU}}(A,B)=
\begin{cases}
1- \frac{\sum_{a \in A \cap B}w(a)}{\sum_{a \in A \cup B} w(a)}, & \textup {if \ }  A \cup B  \neq \emptyset , \\ 
0,   & \textup {otherwise.} 
\end{cases}$$ 
\refstepcounter{myindex} 
\label{ex.J.no.NEU}

\noindent \TR:
define $I_{\ref{ex.J.no.TR2}}$  by
 $$I_{\ref{ex.J.no.TR2}}(A,B)=
\begin{cases}
\frac{2\card{A \setminus B} +  \card{B \setminus A}}{\card{A \cup B} }, & \textup {if \ }  A \cup B  \neq \emptyset , \\ 
0,   & \textup {otherwise.} 
\end{cases}$$
 \refstepcounter{myindex}  
 \label{ex.J.no.TR2}

\noindent \ERJ: 
$-J$ or 
  $I_{\ref{ex.H.d.no.OSE}}$.

\noindent \TES:  
$$I_{\ref{ex.J.no.ES}}(A,B)=
\begin{cases}
1 - \frac{\card{A \cap B}}{\card{A \cup B}}, & \textup {if \ }  A \cup B  \neq \emptyset , \\ 
1,   & \textup {otherwise.} 
\end{cases}$$
 \refstepcounter{myindex}  
 \label{ex.J.no.ES}

\noindent \RI:
$H$.

\noindent \ADD:  $S$.
\epr

\noindent {\bf Proof of Theorem~\ref{theo.char.J.dist2}.} 
The necessity part is easy and is thus omitted.
 Thanks  to Lemma~\ref{lemma.type},   \TSES\ can  be rewritten as
 $I(i,j,i)-I(i,j+1,i-1) = I(i,j+1,i-1)-I(i,j+2,i-2)$.
 Thanks to Theorem~\ref{theo.char.J.ordering}, we know that $I$ is a numerical representation of $\succsim_{J}$. There exists therefore $\phi: \RR \to \RR$, strictly increasing, such that $I = \phi(J)$. Hence, 
$$\phi\left( \frac{i+i}{i+j+i} \right)- \phi\left( \frac{i+i-1}{i+j+i} \right) = \phi\left( \frac{i+i-1}{i+j+i} \right) - \phi\left( \frac{i+i-2}{i+j+i} \right),$$
for all $i,j \in \NN$ such that $i+j>0$. Let us define $z=2i+j$ and 
$\psi_{z}(h)= \phi\big((z-h)/z\big)$. So, $\psi_{z}(j) - \psi_{z}(j+1) = \psi_{z}(j+1) - \psi_{z}(j+2)$. Writing this condition for $j = 0, 1, 2, \ldots$, we find 
\begin{align*}
\psi_{z}(2)  & = 2\psi_{z}(1) - \psi_{z}(0)  = \psi_{z}(0) + 2\big(\psi_{z}(1) - \psi_{z}(0) \big), \\
\psi_{z}(3)  & = 2\psi_{z}(2) - \psi_{z}(1)  = \psi_{z}(0) + 3\big(\psi_{z}(1) - \psi_{z}(0) \big), \\
\psi_{z}(4)  & = 2\psi_{z}(3)  - \psi_{z}(2) =  \psi_{z}(0) + 4\big(\psi_{z}(1) - \psi_{z}(0) \big),\\
\ldots 
\end{align*}
Hence, for all $j \in \{0, 1, \ldots \}$,
$ \phi \left(\frac{z-j}{z}\right) = \psi_{z}(j)= s_{z} + j r_{z} ,$
with $s_{z} = \psi_{z}(0)$  and $r_{z}= \psi_{z}(1)-\psi_{z}(0)$.
Since $J(A,B)=1$ whenever $A \cap B = \emptyset$, we have  $\phi(1)=s_{z}+ 0 r_{z}$. By \LB, 
$\phi(0)=s_{z} + z r_{z}=0$. 
 This implies $s_{z}=\phi(1)$ and $r_{z}=-\phi(1)/z$. Finally, 
$$ \phi \left(\frac{z-j}{z}\right) = \phi(1) - \phi(1) \frac{j}{z},$$
or $\phi(p)=\phi(1) p$ for all  $p$ in $[0,1] \cap \QQ$. Hence, $I=\phi(1)J$.

We now turn to the proof of the logical independence of the conditions.

\noindent \NEU:
$I_{\ref{ex.J.no.NEU}}$. 

\noindent \TR: $I_{\ref{ex.J.no.TR2}}$.

\noindent \ERJ: $-J$ or $I_{\ref{ex.H.d.no.OSE}}$.

\noindent \TES: $I_{\ref{ex.J.no.ES}}$.

\noindent \RI:
$H$.

\noindent \LB:
$$I_{\ref{ex.J.d2.no.LB}}(A,B)=
\begin{cases}
1-\frac{\card{A\cap B}}{2\card{A\cup B}}, & \textup {if \ }  \card{A \cup B} \neq \emptyset, \\ 
0   & \textup {otherwise.} 
\end{cases}
$$
\refstepcounter{myindex} 
\label{ex.J.d2.no.LB}

\noindent \TSES:
$S$.
\epr

\subsection{S\o rensen-Dice}

\noindent {\bf Proof of Theorem~\ref{theo.S.no.add}.} 
 When $A \cup B \neq \emptyset$, the S\o rensen-Dice distance can be written as
\begin{equation}
\label{eq.SO.add}
\frac{\card{A \setminus B} }{\card{A} + \card{B}} + \frac{ \card{B \setminus A}}{\card{A} + \card{B}}.
\end{equation}
It is the only additive decomposition (if we exclude the trivial decomposition). 
For \GADD\ to hold, $S(\kappa(A ,B) , \lambda(A,B))$ must be equal to one of the  terms in the left-hand side of \eqref{eq.SO.add}. We assume without loss of generality 
$$S(\kappa(A ,B) , \lambda(A,B)) = \frac{\card{A \setminus B} }{\card{A} + \card{B}},$$
 which implies
\begin{equation}
\label{eq.SO.GA}
 \frac{ \card{\kappa(A ,B) \setminus \lambda(A,B)}}{ \card{\kappa(A ,B)} + \card{\lambda(A,B)}} + 
\frac{ \card{\lambda(A ,B) \setminus \kappa(A,B)}}{ \card{\kappa(A ,B)} + \card{\lambda(A,B)}}
 = \frac{\card{A \setminus B} }{\card{A} + \card{B}}.
 \end{equation}
Since $\card{A\setminus B}$ cannot be written as a sum, one of the terms on the left-hand side of \eqref{eq.SO.GA} must be zero. We assume without loss of generality the second one is zero, that is $\card{\lambda(A ,B) \setminus \kappa(A,B)}=0$. Hence  $\lambda(A ,B) \subseteq \kappa(A,B)$.
If $\lambda(A ,B) = \emptyset$, then \eqref{eq.SO.GA} implies $\card{\kappa(A,B)} = \card{A} + \card{B}$, which is not possible because there is no set $\kappa(A,B)$ (written in terms of $A,B, \cap, \cup, \setminus$) such that $\card{\kappa(A,B)} = \card{A} + \card{B}$. So we can assume 
$\emptyset \neq \lambda(A ,B) \subseteq \kappa(A,B)$. Since $\card{\kappa(A ,B) \setminus \lambda(A,B)} = \card{A \setminus B}$, three cases are possible.
\begin{itemize}
\item $\kappa(A ,B)= A \cup B$ and $\lambda(A,B)= B$.
\item $\kappa(A ,B)= A $ and $\lambda(A,B)= A \cap B$.
\item $\kappa(A ,B)= A \triangle B $ and $\lambda(A,B)= B \setminus A$.
\end{itemize}
In none of these cases, 
$\card{\kappa(A ,B)} + \card{\lambda(A,B)}  = \card{A} + \card{B}$. So, \eqref{eq.SO.GA} cannot hold with $\kappa(A,B)$ and $\lambda(A,B)$ written in terms of $A,B, \cap, \cup, \setminus$.
\epr

\noindent {\bf Proof of Theorem~\ref{theo.char.S.dist}.} 
The necessity part is easy and is thus omitted.
 Thanks  to Lemma~\ref{lemma.type},   \TSESS\ can  be rewritten as
 $I(i,j,k)-I(i+2,j-1,k) = I(i+2,j-1,k)-I(i+4,j-2,k)$.
 Thanks to Theorems~\ref{theo.char.J.ordering} and~\ref{theo.char.SD.ordering}, we know that $I$ is a numerical representation of $\succsim_{S}$. There exists therefore $\phi: \RR \to \RR$, strictly increasing, such that $I = \phi(S)$. Hence, 
$$\phi\left( \frac{2j}{i+2j+k} \right)- \phi\left( \frac{2j-2}{i+2j+k} \right) = \phi\left( \frac{2j-2}{i+2j+k} \right) - \phi\left( \frac{2j-4}{i+2j+k} \right),$$
for all $i,j,k \in \NN$ such that $i+2j+k>0$. Let us define $m=2j$, $z=i+2j+k$ and 
$\psi_{z}(h)= \phi(h/z)$. So, $\psi_{z}(m) - \psi_{z}(m-1) = \psi_{z}(m-1) - \psi_{z}(m-2)$. Writing this condition for $m = 2, \ldots, z$, we find 
\begin{align*}
\psi_{z}(2)  & = \psi_{z}(1) + \big(\psi_{z}(1) - \psi_{z}(0) \big) = \psi_{z}(0) + 2 \big(\psi_{z}(1) - \psi_{z}(0) \big), \\
\psi_{z}(3)  & = \psi_{z}(2) + \big(\psi_{z}(1) - \psi_{z}(0) \big) = \psi_{z}(0) + 3 \big(\psi_{z}(1) - \psi_{z}(0) \big), \\
\psi_{z}(4)  & = \psi_{z}(3) + \big(\psi_{z}(1) - \psi_{z}(0) \big) =  \psi_{z}(0) + 4 \big(\psi_{z}(1) - \psi_{z}(0) \big),\\
\ldots \\
\psi_{z}(m)  & = \psi_{z}(m-1) + \big(\psi_{z}(1) - \psi_{z}(0) \big)=   \psi_{z}(0) + m \big(\psi_{z}(1) - \psi_{z}(0) \big) \\
\ldots
\end{align*}
The rest of the proof is similar to that of Theorem~\ref{theo.char.J.dist2}.
We now turn to the proof of the logical independence of the axioms.

\noindent \NEU:
let $w:X \to \NN$  be a a non-constant mapping and 
$$I_{\ref{ex.S.ind.no.NEU}}(A,B)=
 \begin{cases}
1- \frac{ 2 \ \sum_{a \in A \cap B}w(a)}{\sum_{a \in A}w(a)+ \sum_{a \in B}w(a)}, & \textup {if \ }  A \cup B  \neq \emptyset , \\ 
0,   & \textup {otherwise.} 
\end{cases}$$
\refstepcounter{myindex} 
\label{ex.S.ind.no.NEU}

\noindent \TR:
$$I_{\ref{ex.S.ind.no.TR}}(A,B)= 
\begin{cases}
1 - \frac{\card{A \cap B} \min(\card{A}, \card{B})}{\card{A}^{2}+ \card{B}^{2}}, & \textup {if \ }  A \neq \emptyset \neq B, \\ 
0 , & \textup {if \ }  A = \emptyset = B, \\ 
1   & \textup {otherwise.} 
\end{cases}
$$
\refstepcounter{myindex} \label{ex.S.ind.no.TR}

\noindent \ERJ:
$-S$ or $I_{\ref{ex.H.d.no.OSE}}$. 

\noindent \TES: 
$$I_{\ref{ex.S.ind.no.TES}}(A,B)= 
\begin{cases}
1 - \frac{2\card{A \cap B} }{\card{A}+ \card{B}}, & \textup {if \ }  A \neq \emptyset \neq B, \\ 
0 , & \textup {if \ }  A = \emptyset = B, \\ 
1   & \textup {otherwise.} 
\end{cases}
$$
\refstepcounter{myindex} \label{ex.S.ind.no.TES}

\noindent \RI:  $H$.

\noindent \TSESS: $J$.

\noindent \LB: $1+S$. 
\epr

\subsection{Overlap}

\noindent {\bf Proof of Theorem~\ref{theo.char.O.ordering}.}
Consider any $A,B,C,D \in Y$ with types $x=t^{AB}$ and $y = t^{CD}$ and suppose without loss of generality $O(A,B) \geq O(C,D)$.  We must show that $(A,B) \succsim (C,D)$. 

Let us first consider the cases where $(A,B)$ or $(C,D)$ is equal to $(\emptyset,\emptyset)$. 
\begin{itemize}
\item If $(C,D)=(\emptyset,\emptyset)=(A,B)$, then obviously $(A,B) \sim (C,D)$. 

\item  If $(C,D)=(\emptyset,\emptyset)\neq (A,B)$, then, by \TES, $(C,D) \sim (\{a \},\{ a \}) \sim (0,1,0)$,  with $a$ as in the statement of \TES. By \RI, $(0,1,0) \sim (0,x_{2},0)$. By \EI, $(0,x_{2},0) \sim (x_{1},x_{2},0)$. By \ERO\ \partone,  $(x_{1},x_{2},0) \precsim (x_{1},x_{2},x_{3})\sim (A,B)$ (the comparison is not strict because $x_{3}$ can be zero).
By transitivity, $(C,D) \precsim (A,B)$.

\item  If $(A,B)=(\emptyset,\emptyset)\neq (C,D)$, then $O(A,B) \geq O(C,D)$ implies $O(C,D)=0$ and, hence, $C\subseteq D$ or $D \subseteq C$. By \SYM, we can assume without loss of generality $D \subseteq C$, which implies $(C,D) \sim (y_{1},y_{2},0)$.
Then, by \TES, 
$(A,B) = (\emptyset,\emptyset) \sim (0,0,0) \sim (0,1,0)$. By \RI, $(0,1,0) \sim (0,y_{2},0)$. By \EI, $(0,y_{2},0) \sim (y_{1},y_{2},0)$.
By transitivity, $(A,B) \sim (C,D)$.

\end{itemize}
The rest of the proof assumes $(A,B) \neq (\emptyset,\emptyset) \neq (C,D)$. Let us now consider the cases where exactly one of $A,B$ or exactly one of $C,D$ is empty.
\begin{itemize}
\item Suppose $A=\emptyset, B \neq \emptyset, C \neq \emptyset \neq D$. By \RI, $(A,B) \sim (x_{1},0,0) \sim (1,0,0)$. 
\begin{itemize}
\item Suppose $y_{3}>0$. \OES\ and \RI\ imply $(1,0,0) \sim (1,0,1) \sim (y_{3},0,y_{3})$. By \ERO\ \parttwo, $(y_{3},0,y_{3}) \succsim (y_{3},y_{2},y_{3})$ (the comparison is not strict because $y_{2}$ can be zero). By \EI, $(y_{3},y_{2},y_{3}) \sim (y_{1},y_{2},y_{3})$. By transitivity, $(A,B) \succsim (C,D)$.
\item Suppose $y_{3}=0$.  \RI\  and \ERO\ \parttwo\ imply $(1,0,0) \sim (y_{1},0,0) \succsim (y_{1}, y_{2},0)$ (the comparison is not strict because $y_{2}$ can be zero). By transitivity, $(A,B) \succsim (C,D)$.
\end{itemize}

\item Suppose $A\neq \emptyset, B = \emptyset, C \neq \emptyset \neq D$. By \SYM, this case is equivalent to the previous one.

\item Suppose $A \neq \emptyset \neq B, C = \emptyset, D \neq \emptyset$. Then $O(C,D)=1$ and $O(A,B) \geq O(C,D)$ imply $O(A,B)=1$. Since $(A,B) \neq (\emptyset,\emptyset)$, $A \cap B = \emptyset$ must hold. So, $(A,B) \sim (x_{1}, 0, x_{3})$ and $(C,D) \sim (y_{1},0,0)$. By \RI, $(y_{1},0,0) \sim (1,0,0)$. By \OES, $(1,0,0) \sim (1,0,1)$. By \RI, $(1,0,1) \sim (x_{3},0,x_{3})$. By \EI, $(x_{3},0,x_{3}) \sim (x_{1},0,x_{3})\sim (A,B)$. Transitivity concludes.

\item Suppose $A \neq \emptyset \neq B, C \neq \emptyset, D = \emptyset$. By \SYM, this case is equivalent to the previous one.

\item Suppose $A = \emptyset, B \neq \emptyset, C = \emptyset, D \neq \emptyset$ (or one of the three other cases that are equivalent by \SYM). By \RI, $(A,B) \sim (x_{1},0,0) \sim (y_{1},0,0) \sim (C,D)$. Transitivity concludes.

\end{itemize}
The rest of the proof assumes none of  $A,B,C,D$ is empty.

Note that, when $x_{2}, y_{2} >0$, we have $O(A,B) \geq O(C,D)$ iff $x_{2}/(x_{2}+x_{3}) \leq y_{2}/(y_{2}+y_{3})$ iff $x_{2}(y_{2}+y_{3}) \leq  y_{2} (x_{2}+x_{3})$ iff $x_{3}(y_{2}+y_{3}) \geq  y_{3} (x_{2}+x_{3})$. Note also that $y_{2} (x_{2}+x_{3})-x_{2}(y_{2}+y_{3}) = y_{2}x_{3}-x_{2}y_{3}$.

\begin{itemize}

\item If $O(A,B) = O(C,D)=0$, then $(A,B) \sim (x_{1},x_{2},0)$ and $(C,D) \sim (y_{1},y_{2},0)$. By \EI\ and \RI, $(x_{1},x_{2},0) \sim (x_{2},x_{2},0)  \sim (y_{2},y_{2},0)$. By \EI, $(y_{2},y_{2},0) \sim (y_{1},y_{2},0)$. By transitivity, $(A,B) \sim (C,D)$.

\item If $1>O(A,B) = O(C,D) >0$, then \RI\ and \EI\  imply $x \sim  \big(x_{1} (y_{2}+y_{3})(x_{2}+x_{3}),x_{2} (y_{2}+y_{3}),x_{3} (y_{2}+y_{3})\big) = \big(x_{1} (y_{2}+y_{3})(x_{2}+x_{3}), y_{2} (x_{2}+x_{3}), y_{3} (x_{2}+x_{3})\big) \sim \big(x_{1} (y_{2}+y_{3}), y_{2} , y_{3} \big)$. 
By transitivity, $x \sim  \big(x_{1} (y_{2}+y_{3}), y_{2} , y_{3} \big)$. 
Since $x_{1} (y_{2}+y_{3}) > y_{3}$ and $y_{1}>y_{3}$,  \EI\  implies $x \sim (y_{1},y_{2},y_{3})$. Notice that the type $ \big(x_{1} (y_{2}+y_{3})(x_{2}+x_{3}),x_{2} (y_{2}+y_{3}),x_{3} (y_{2}+y_{3})\big)$ corresponds to an existing pair of sets because we have assumed $X$ is infinite.

\item If $1=O(A,B) = O(C,D)$, then $x_{2}=y_{2}=0$. By \EI, $x = (x_{1},0,x_{3}) \sim (x_{3},0,x_{3})$. By \RI, $(x_{3},0,x_{3}) \sim (y_{3},0,y_{3})$. By \EI, $(y_{3},0,y_{3}) \sim (y_{1},0,y_{3})=y$. By transitivity, all these equivalences imply $x \sim y$.

\item If $1 > O(A,B) > O(C,D) \geq 0$,  as previously, $x \sim  \big(x_{1} (y_{2}+y_{3})(x_{2}+x_{3}),x_{2} (y_{2}+y_{3}),x_{3} (y_{2}+y_{3})\big)$. 
By \ERO\ \parttwo, $\big(x_{1} (y_{2}+y_{3})(x_{2}+x_{3}),x_{2} (y_{2}+y_{3}),x_{3} (y_{2}+y_{3})\big) \succ \big(x_{1} (y_{2}+y_{3})(x_{2}+x_{3}),y_{2} (x_{2}+x_{3}),x_{3} (y_{2}+y_{3})\big)$. 
By \ERO\ \partone, $\big(x_{1} (y_{2}+y_{3})(x_{2}+x_{3}),y_{2} (x_{2}+x_{3}),x_{3} (y_{2}+y_{3})\big) \succ \big(x_{1} (y_{2}+y_{3})(x_{2}+x_{3}),y_{2} (x_{2}+x_{3}),y_{3} (x_{2}+x_{3})\big)$. By \RI, $\big(x_{1} (y_{2}+y_{3})(x_{2}+x_{3}),y_{2} (x_{2}+x_{3}),y_{3} (x_{2}+x_{3})\big) \sim \big(x_{1} (y_{2}+y_{3}),y_{2} ,y_{3} \big)$. \EI\  then implies $\big(x_{1} (y_{2}+y_{3}),y_{2} ,y_{3} \big) \sim (y_{1},y_{2},y_{3})$. Using transitivity, we can chain all these comparisons to obtain 
$x \succ (y_{1},y_{2},y_{3})$.

\item If $1 = O(A,B) > O(C,D) > 0$, then $y_{2} > x_{2} =0$. As previously, $x = (x_{1},0,x_{3}) \sim (x_{3},0,x_{3}) \sim (y_{3},0,y_{3}) \sim (y_{1},0,y_{3})$. By \ERO\ \parttwo, $(y_{1},0,y_{3}) \succ (y_{1},y_{2},y_{3})=y$. By transitivity, $x \succ y$.

\item If $1 = O(A,B) > O(C,D) = 0$, then $y_{2} > x_{2} =0$ and $y_{3}=0$. By \ERO\ \partone\ and \RI, $(A,B) \sim (x_{1},0,x_{3}) \succ (x_{1},0,0) \sim (y_{1},0,0)$. By \ERO\ \parttwo, $(y_{1},0,0) \succ (y_{1},y_{2},0)$. By transitivity, $(A,B) \succ (C,D)$.

\end{itemize}
The logical independence of the axioms of Theorem~\ref{theo.char.O.ordering} will be established in the proof of Theorem~\ref{theo.char.O.dist} about the Overlap distance.
\epr

\noindent {\bf Proof of Theorem~\ref{theo.O.no.add}.}
Similar to that of Theorem~\ref{theo.S.no.add}.

\medskip

\noindent {\bf Proof of Theorem~\ref{theo.char.O.dist}.}
The necessity part is easy and is thus omitted. 
Thanks  to Lemma~\ref{lemma.type},   \TSES\ can  be rewritten as
$I(i,j,k)-I(i,j+1,k-1) = I(i,j+1,k-1)-I(i,j+2,k-2)$. 
Thanks to Theorem~\ref{theo.char.O.ordering}, we know that $I$ is a numerical representation of $\succsim_{O}$. There exists therefore $\phi: \RR \to \RR$, strictly increasing, such that $I = \phi(O)$. Hence, 
$$\phi\left( \frac{j}{j+k} \right)- \phi\left( \frac{j+1}{j+k} \right) = \phi\left( \frac{j+1}{j+k} \right) - \phi\left( \frac{j+2}{j+k} \right),$$
for all $j,k \in \NN$. The rest of the proof is similar to that of Theorems~\ref{theo.char.J.dist2} and  \ref{theo.char.S.dist}.

We now prove the logical independence of the axioms.

\noindent \NEU: 
let $w:X \to \NN$ be a non-constant mapping and 
$$I_{\ref{ex.O.no.NEU}}(A,B)=
\begin{cases}
1-\frac{\sum_{a \in A \cap B}w(a)}{\min\left(\sum_{a \in A}w(a), \sum_{a \in B}w(a)\right)}, & \textup {if \ }  A   \neq \emptyset \neq B , \\ 
0, & \textup {if \ }  A   = \emptyset = B , \\ 
1,   & \textup {otherwise.} 
\end{cases}$$ 
\refstepcounter{myindex} 
\label{ex.O.no.NEU}

\noindent \SYM:
$$I_{\ref{ex.O.no.SYM}}(A,B)=
\begin{cases}
\frac{ \card{A \triangle B} - \max(\card{A} - \card{B},0)}{\card{A \cup B} - \max(\card{A} - \card{B},0)}, & \textup {if \ }  A \neq \emptyset \neq B, \\ 
0, & \textup {if \ }  A = \emptyset = B, \\ 
1,   & \textup {otherwise.} 
\end{cases}
$$
\refstepcounter{myindex} 
\label{ex.O.no.SYM}

\noindent \RI:
$$I_{\ref{ex.O.no.RI}}(A,B)=
\begin{cases}
 1 - \frac{\card{A \cap B}}{1+\min(\card{A} , \card{B})}, &\textrm{if} A \setminus B \neq \emptyset \neq B \setminus A, \\
 1, &\textrm{if exactly one of } A,B   \textrm{ is empty, } \\
 0, &\textrm{otherwise}.
\end{cases}
$$
\refstepcounter{myindex} 
\label{ex.O.no.RI}

\noindent \OES:
$$I_{\ref{ex.O.no.OES}}(A,B)=
\begin{cases}
1 - \frac{\card{A \cap B}}{\min(\card{A} , \card{ B})}, & \textup {if \ }  A  \neq \emptyset \neq B, \\ 
0,   & \textup {if \ }  A  = \emptyset  = B \\
1/2,   & \textup {otherwise.}  
\end{cases}
$$
\refstepcounter{myindex} 
\label{ex.O.no.OES}

\noindent \TES:
$$I_{\ref{ex.O.no.TES}}(A,B)=
\begin{cases}
1 - \frac{\card{A \cap B}}{\min(\card{A} , \card{ B})}, & \textup {if \ }  A  \neq \emptyset \neq B, \\ 
1,   & \textup {if \ }  A  = \emptyset  = B \\
1,   & \textup {otherwise.}  
\end{cases}
$$
\refstepcounter{myindex} 
\label{ex.O.no.TES}

\noindent \EI:
$J$.

\noindent \ERO:
$I_{\ref{ex.H.d.no.OSE}}$. 

\noindent \LB:
$1+O.$

\noindent \TSES:
$O^{2}.$
\epr

\section*{Data availability}

We do not analyse or generate any datasets, because our work proceeds within a theoretical and mathematical approach.

\bibliography{mybib_Sandip}

\backmatter
\bmhead{Acknowledgments}
Research on this project  started when Thierry Marchant was visiting  BITS Pilani, K K Birla Goa Campus, India in January 2025. Thierry Marchant thanks Snehanshu Saha for supportive facilities at the BITS Goa campus. Sandip Sarkar acknowledges  ANRF for funding from ANRF-MATRICS project number (ANRF/ARGM/2025/000737/QSS).
We also thank Denis Bouyssou, Georgios Gerasimou, Bj\o rn Kjos-Hanssen, and Marc Pirlot for fruitful discussions.

\end{document}